\documentclass[aps,showpacs,preprintnumbers,amsmath, amssymb]{revtex4}

\usepackage{float}
\usepackage{graphics,epsfig}
\usepackage{graphicx}
\usepackage{dcolumn}
\usepackage{bm}

\begin{document}
\baselineskip=0.8 cm

\title{{\bf Bounds on the minimum orbital period of general static
spherical black holes: \\ a sharp upper bound and a conditional lower
bound}}

\author{Yan Peng$^{1,2}$\footnote{Corresponding author: yanpengphy@163.com}}
\author{Guohua Liu$^{1}$\footnote{Corresponding author: liuguohua1234@163.com}}
\affiliation{\\$^{1}$ College of Mathematics and Physics, Xinjiang Agricultural University, Urumqi, Xinjiang, 830052,P.R.China}
\affiliation{\\$^{2}$ School of Mathematical Sciences, Qufu Normal University, Qufu, Shandong 273165,P.R.China}

\pacs{04.70.-s, 04.20.-q, 97.60.Lf}

\begin{abstract}
We investigate bounds on the minimum orbital period around four-dimensional
general static spherical black holes. For all static spherically symmetric
black holes with an event horizon, asymptotic flatness and non-negative
energy density outside the horizon, we prove $T_{\min}\leq 6\sqrt{3}\,\pi M$.
The bound is sharp and is saturated by the Schwarzschild solution.
The lower bound requires an additional condition on the exterior matter:
under the tangential pressure condition
$p_{\tau}\geq(\kappa/2)\rho$ with $\kappa>1$, imposed for all $r\geq r_{g}$,
where $r_{g}$ is the largest radius at which $T(r)$ attains its global
minimum, we prove $T_{\min}\geq c(\kappa)M$ with
$c(\kappa)=6\sqrt{3}\pi(\kappa-1)\sqrt{\kappa+2}/\kappa^{3/2}$.
In summary, both bounds apply to the same geometric class and differ only in
what they demand of the matter: non-negative energy density suffices for the
upper bound, while the lower bound is established under the additional
tangential pressure condition.
\end{abstract}

\maketitle

\section{Introduction}

The motion of test particles and light rays around black holes is one of
the oldest and most informative probes of strong-field gravity
\cite{KS1}-\cite{KS11}. Of particular interest is the shortest possible
orbital period \cite{Hod1,Hod2,peng1,PY}. For lightlike circular orbits,
the orbital period measured by asymptotic observers is $T=2\pi/\Omega$,
where $\Omega$ is the angular velocity of the orbit. In the static
spherical setting considered below, the orbital-period function diverges
both at the horizon and at infinity, so its global minimum is attained
at a photon sphere. The minimum orbital period
$T_{\min}$ of lightlike circular orbits around four-dimensional black
holes has been conjectured to obey the universal bounds
$4\pi M \le T_{\min}\le 6\sqrt{3}\,\pi M$,
where $M$ is the black hole mass \cite{peng2,peng3}. Whether these bounds
hold universally is an interesting question to be explored.

Consider first the upper bound. Model-by-model studies indicate $T_{\min}\le 6\sqrt{3}\,\pi M$
for Schwarzschild black holes, Reissner--Nordstr\"om, Kerr-Newman, and the non-singular Hayward
and Bardeen black holes \cite{peng2,peng3,Hayward,Bardeen1,Bardeen2,Bardeen3}.
The upper bound is saturated in the Schwarzschild limit in every case. All of
these results are either restricted to specific solutions or supported numerically.
In this paper, we prove that the upper bound is a theorem for the entire class of
static spherically symmetric black holes with asymptotic flatness and non-negative energy density outside the horizon.
The Schwarzschild, Reissner--Nordstr\"om, Hayward and Bardeen
solutions are covered by this general result.

The lower bound has a different status. The $4\pi M$ bound is a theorem
for the rotating Kerr \cite{Hod1} and Kerr--Newman families \cite{peng1},
but these estimates cannot be extended directly to the general static
class. For static black holes, the minimum period is controlled by the
local compactness at a photon sphere, while the total ADM mass also
includes contributions from matter arbitrarily far away. We construct
asymptotically flat spacetimes satisfying the weak, null and strong
energy conditions, for which $T_{\min}/M$ can be arbitrarily small.
Thus, when $M$ is the total ADM mass, these energy conditions alone
cannot give a universal positive lower bound. The two bounds therefore
require different conditions on the matter: the upper bound follows
from the monotonicity and boundedness of the mass function, whereas
the lower bound requires a relation between the local photon-sphere
compactness $\mu$ and the total mass $M$. An additional condition on
the exterior matter distribution is needed to establish such a relation.
We will show that a one-parameter tangential-pressure condition provides
the required control.

The paper is organized as follows. Section II sets up the general static
spherical framework and proves the upper bound theorem. Section III
presents the energy conditions, explains why additional control of the
exterior matter is necessary, and establishes the lower bound under the
tangential-pressure condition. The standard black hole models are also
examined in this section. Section IV translates the bounds into a
statement about light-ring frequencies, and Section V concludes with
open problems.

\section{An analytic upper bound for general static spherical black holes}
\label{sec:upper}

We consider an important subclass of static spherically symmetric
spacetimes with $g_{tt}g_{rr}=-1$. This class contains the
Schwarzschild and Reissner--Nordstr\"om solutions and a wide range
of regular black holes.
The metric can be written as \cite{Hayward,ZYX}
\begin{equation}\label{eq:metric-general}
ds^{2}=-f(r)\,dt^{2}+\frac{dr^{2}}{f(r)}+r^{2}\!\left(d\theta^{2}
+\sin^{2}\theta\,d\phi^{2}\right),
\end{equation}
where $f(r)=1-\frac{2m(r)}{r}$ and $m(r)$ is the mass enclosed within
the sphere of radius $r$. We assume that $m(r)$ is continuously
differentiable outside the horizon, as required in the proof of the
upper bound. For the lower-bound analysis in Section III, we further
assume that $m(r)$ is twice continuously differentiable for
$r\geq r_{g}$. These are regularity assumptions on the mass function,
rather than additional conditions on the matter. We assume:

\begin{itemize}
\item[(i)] \emph{Asymptotic flatness:} $m(r)\to M$ as $r\to\infty$, where
$M$ is the total ADM mass.
\item[(ii)] \emph{Non-negative energy density:} the local energy density
measured by a static observer outside the horizon is non-negative
\cite{Hod3}. For the mass function used here, this density is written as
$\rho(r)=m'(r)/(4\pi r^{2})$, so the assumption is equivalent to
$m'(r)\geq0$ for all $r$ outside the horizon.
So this assumption is equivalent to $m'(r)\geq0$ for all $r$ outside the
horizon. We do not require the full weak energy condition for the upper-bound
theorem.
\item[(iii)] \emph{Existence of an event horizon} at $r=r_{h}$,
where $r_{h}$ is the largest root of $f(r)=0$. Thus $f(r)>0$ for all
$r>r_{h}$, and $m(r_{h})=r_{h}/2$.
\end{itemize}

For a circular null curve in the equatorial plane
($ds^{2}=dr=d\theta=0$, $\theta=\pi/2$), the metric
\eqref{eq:metric-general} gives $f(r)\,dt^{2}=r^{2}d\phi^{2}$.
Thus the coordinate angular velocity is
$\Omega(r)=d\phi/dt=\sqrt{f(r)}/r$ \cite{KS9}, and the corresponding
period is
\begin{equation}\label{eq:T-general}
T(r)=\frac{2\pi}{\Omega(r)}=\frac{2\pi r}{\sqrt{f(r)}}\,.
\end{equation}
At a critical point of $T(r)$, the circular null curve is a null
geodesic, and $T(r)$ gives the orbital period measured by an asymptotic
observer.

Differentiating \eqref{eq:T-general} with respect to
$r$, we obtain
\begin{equation}\label{eq:Tprime-general}
T'(r)=2\pi\,\frac{f(r)-\tfrac{r}{2}f'(r)}{f(r)^{3/2}}.
\end{equation}
Using $f=1-2m/r$ and $f'=2m/r^{2}-2m'/r$, the numerator simplifies to
\begin{equation}\label{eq:key}
f(r)-\frac{r}{2}f'(r)
=1-\frac{3m(r)}{r}+m'(r).
\end{equation}

So the critical points of $T(r)$ satisfying
$T'(r)=0$ are determined by
\begin{equation}\label{eq:critical}
m'(r)=\frac{3m(r)}{r}-1,
\end{equation}
which is precisely the photon-sphere condition $f(r)-\frac{r}{2}f'(r)=0$.
Note that $T(r)\to+\infty$ both at the horizon (since $f\to0^{+}$) and
at infinity (since $T\sim2\pi r$). Therefore $T$ must attain its
minimum $T_{\min}$ at an interior critical point. There may be several
critical points. Let $u$ be the outermost one. It need not be the point
where $T$ attains its global minimum, so in general
$T_{\min}\leq T(u)$. If $T'(r_0)<0$ at some $r_0>u$, then the subsequent
divergence $T(r)\to+\infty$ would require another critical point farther
out, contradicting the definition of $u$. Consequently,
\begin{equation}\label{eq:Tprime-positive}
T'(r)\geq0\quad\text{for all }r>u,
\end{equation}
which by \eqref{eq:Tprime-general}--\eqref{eq:key} is equivalent to
\begin{equation}\label{eq:mprime-ineq}
m'(r)\geq\frac{3m(r)}{r}-1\qquad\text{for all }r>u.
\end{equation}
We will prove $T(u)\le6\sqrt{3}\pi M$, and the bound for the global minimum
follows from $T_{\min}\le T(u)$.

Evaluating \eqref{eq:critical} at $u$ and using the non-negative energy
density condition $m'\geq0$ gives
\begin{equation}\label{eq:v-lower}
m(u)\geq\frac{u}{3}.
\end{equation}
On the other hand, $m(u)\leq M$, hence
\begin{equation}\label{eq:v-gx}
u\leq3m(u)\leq3M.
\end{equation}

It follows from \eqref{eq:v-lower} that there are two cases to consider:
\begin{equation}
m(u)=\frac{u}{3},
\qquad\text{or}\qquad
m(u)>\frac{u}{3}.
\end{equation}
We now prove the upper bound in each case.

\smallskip
\noindent\emph{Case 1: $m(u)=u/3$.} In this case,
$f(u)=1-2m(u)/u=1/3$, and using \eqref{eq:T-general} and $T_{\min}\leq T(u)$,
we obtain
\begin{equation}
T_{\min}\leq T(u)=\frac{2\pi u}{\sqrt{1/3}}
=2\sqrt{3}\,\pi u=6\sqrt{3}\pi m(u)
\leq6\sqrt{3}\pi M.
\end{equation}

\smallskip
\noindent\emph{Case 2: $m(u)>u/3$.} In this case, the metric function satisfies
\begin{equation}
f(u)<1/3,
\end{equation}
and we define
\begin{equation}
S=\Bigl\{r\geq u:\ r\geq3m(r)\Bigr\}
=\Bigl\{r\geq u:\ f(r)\geq\frac{1}{3}\Bigr\}.
\end{equation}
The set $S$ consists of the radii beyond $u$ at which $f(r)$ has reached
or exceeded $1/3$.

Since $f(u)<1/3$, the set $S$ consists of the radii beyond $u$ at which
$f(r)$ has reached or exceeded $1/3$. The point of introducing $S$ is
that its first point $r_{f}=\inf S$ marks the largest interval
$[u,r_{f})$. On $[u,r_{f})$ we have $f<1/3$, so \eqref{eq:mprime-ineq} forces $m$
to grow at the controlled rate $m'(r)\geq\frac{3m(r)}{r}-1>0$,
and this inequality can be integrated up to $r_{f}$. At $r_{f}$
the value $m(r_{f})=r_{f}/3$ is fixed exactly by $f(r_{f})=1/3$,
and the mass budget $m\leq M$ then caps the exit point at $r_{f}\leq 3M$.
In this way the local quantity $T(u)$ is converted into a bound on the ADM mass.

The set $S$ is non-empty: for any $r>3M$, we have
$3m(r)\leq3M<r$. Let $r_f=\inf S$. Since $f(u)<1/3$, continuity gives
$u<r_f\leq3M$ and
\begin{equation}\label{eq:rf}
m(r_f)=\frac{r_f}{3},
\end{equation}
or equivalently $f(r_f)=1/3$.

By definition of $r_f$, we have $r<3m(r)$ throughout $[u,r_f)$.
Thus \eqref{eq:mprime-ineq} gives a strict differential inequality
on this interval:
\begin{equation}\label{eq:ODE}
m'(r)\geq\frac{3m(r)}{r}-1>0,\qquad u\leq r<r_f.
\end{equation}

Dividing \eqref{eq:ODE} by $r^{3}$, we obtain
\begin{equation}
\frac{m'(r)}{r^{3}}-\frac{3m(r)}{r^{4}}
\geq-\frac{1}{r^{3}},\qquad u\leq r<r_f.
\end{equation}
The left-hand side is the derivative of $m(r)/r^{3}$ with
respect to $r$: $\frac{d}{dr}\left(\frac{m(r)}{r^{3}}\right)
=\frac{m'(r)}{r^{3}}-\frac{3m(r)}{r^{4}}$.
Thus, the above inequality can be written as
\begin{equation}
\frac{d}{dr}\left(\frac{m(r)}{r^{3}}\right)
\geq-\frac{1}{r^{3}},\qquad u\leq r<r_f.
\end{equation}
We integrate this inequality from $u$ to $r$ ($u\leq r\leq r_{f}$),
\begin{equation}
\frac{m(r)}{r^{3}}-\frac{m(u)}{u^{3}}
\geq\int_{u}^{r}\frac{ds}{-s^{3}}
=\frac{1}{2r^{2}}-\frac{1}{2u^{2}}.
\end{equation}
We further obtain an estimate for $m(r)$:
\begin{equation}\label{eq:gronwall}
m(r)\geq r^{3}\!\left(\frac{m(u)}{u^{3}}-\frac{1}{2u^{2}}\right)+\frac{r}{2}.
\end{equation}

We set $r=r_f$ in \eqref{eq:gronwall} and use $m(r_f)=r_f/3$ \eqref{eq:rf} to obtain
\begin{equation}
\frac{r_f}{3}\geq r_f^{3}
\left(\frac{m(u)}{u^{3}}-\frac{1}{2u^{2}}\right)
+\frac{r_f}{2}.
\end{equation}
We then rearrange the terms to obtain
\begin{equation}
r_f^{2}\left(\frac{1}{2u^{2}}-\frac{m(u)}{u^{3}}\right)
\geq\frac{1}{6}.
\label{eq:rfbound}
\end{equation}

Since $u$ lies outside the horizon, we have $f(u)>0$, and therefore
$m(u)<u/2$. It follows that
\begin{equation}
\frac{1}{2u^{2}}-\frac{m(u)}{u^{3}}
=\frac{u-2m(u)}{2u^{3}}>0.
\end{equation}
We can thus divide \eqref{eq:rfbound} by this positive coefficient and obtain
\begin{equation}
r_f^{2}\geq\frac{u^{3}}{3(u-2m(u))}.
\label{eq:rfu}
\end{equation}

From \eqref{eq:rf}, we have
$r_f/3=m(r_f)\leq M$, and therefore $r_f\leq3M$.
Combining this upper bound for $r_f$ with \eqref{eq:rfu}, we obtain
\begin{equation}
\frac{u^{3}}{3(u-2m(u))}\leq9M^{2}
\quad\Longrightarrow\quad
\frac{u^{3}}{u-2m(u)}\leq27M^{2}.
\end{equation}

Then we obtain the upper bound on the minimum orbital period:
\begin{equation}
T_{\min}\leq T(u)=\frac{2\pi u}{\sqrt{1-2m(u)/u}}
=2\pi\sqrt{\frac{u^{3}}{u-2m(u)}}
\leq2\pi\sqrt{27}\,M=6\sqrt{3}\pi M.
\end{equation}
The proof uses only asymptotic flatness, the non-negative energy density
condition and the existence of the horizon.
This bound is sharp and is saturated by the Schwarzschild solution.

The above two cases complete the proof of the upper bound. For $m(u)=\frac{u}{3}$, we
obtained $T(u)=6\sqrt{3}\pi m(u)\leq6\sqrt{3}\pi M$, and for
$m(u)>\frac{u}{3}$ the chain through $r_{f}$ gave
$T(u)\leq2\pi\sqrt{27}\,M=6\sqrt{3}\pi M$. Hence
$T_{\min}\leq6\sqrt{3}\pi M$ in either case, and the upper bound is proved.
The Reissner--Nordstr\"om black holes with $m(r)=M-\frac{Q^{2}}{2r}$,
Hayward black holes with $m(r)=\frac{Mr^{3}}{r^{3}+2L^{2}M}$,
and Bardeen black holes with $m(r)=\frac{Mr^{3}}{(r^{2}+g^{2})^{3/2}}$
\cite{peng2} satisfy assumptions \textup{(i)--(iii)} and are covered by
this theorem.
Fig.1 shows the ratio $T_{\min}/M$ over the full parameter range of
the standard models. For each model, $T_{\min}/M$ decreases
monotonically as the parameter increases, and the upper bound is
reached only in the Schwarzschild limit.

\begin{figure}
\centering
\includegraphics[width=0.48\textwidth]{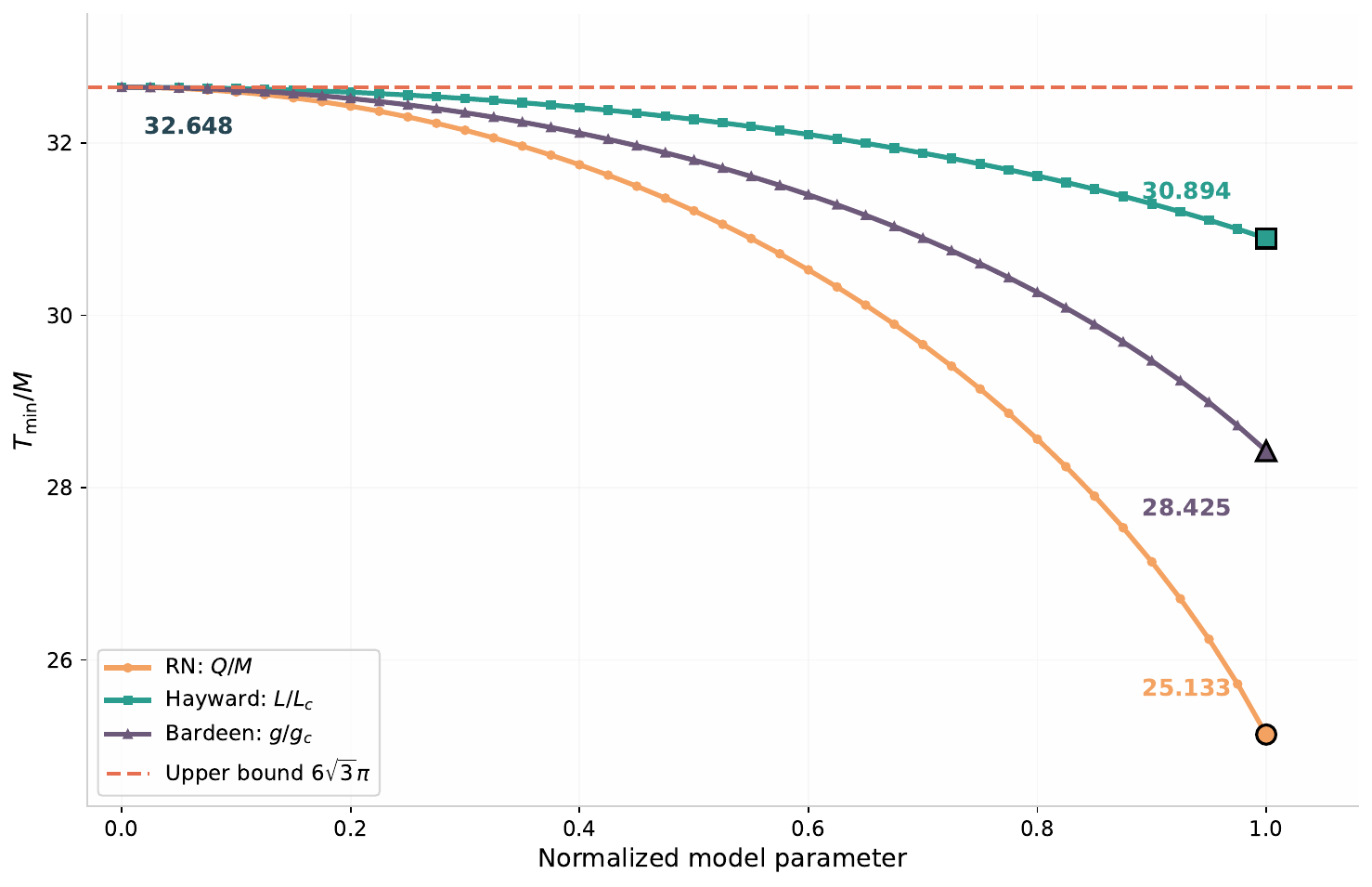}
\caption{The ratio $T_{\min}/M$ for the Reissner--Nordstr\"om ($Q/M$), Hayward ($L/L_{c}$) and Bardeen ($g/g_{c}$) families, as a function of the normalized parameter. In each family $T_{\min}/M$ decreases monotonically from the Schwarzschild value $6\sqrt{3}\pi$ (dashed line) at zero parameter to its minimum at the extremal value of the parameter; the extremal values agree with Table~I. Every point lies below the upper bound, with equality attained only in the Schwarzschild limit.}
\label{fig:scan}
\end{figure}

\section{A lower bound under tangential-pressure conditions}
\label{sec:lower}

For the lower bound, additional
control of the exterior matter is necessary. We first express the energy
conditions in terms of the mass function, then show why the standard energy
conditions alone cannot give a universal positive lower bound in terms of the
total ADM mass.
From the metric \eqref{eq:metric-general}, the mixed components of the
Einstein tensor are
\begin{equation}\label{G-components}
G^{t}{}_{t}=G^{r}{}_{r}=-\frac{2m'}{r^{2}},\qquad
G^{\theta}{}_{\theta}=G^{\phi}{}_{\phi}=-\frac{m''}{r},
\end{equation}
where primes denote $d/dr$. With $G_{\mu\nu}=8\pi T_{\mu\nu}$ for a
static anisotropic fluid ($T^{t}{}_{t}=-\rho$, $T^{r}{}_{r}=p_r$,
$T^{\theta}{}_{\theta}=T^{\phi}{}_{\phi}=p_\tau$, the subscript
$\tau$ denoting the common tangential pressure), the Einstein
equations give \cite{SFH}
\begin{equation}\label{eq:fluid}
\rho=\frac{m'}{4\pi r^{2}},\qquad p_r=-\rho,\qquad p_\tau=-\frac{m''}{8\pi r}.
\end{equation}
Here the continuous differentiability of $m(r)$ outside the horizon
ensures that $\rho$ is continuous, while the additional assumption that
$m(r)$ is twice continuously differentiable for $r\geq r_g$ ensures that
$p_\tau$ is well defined and continuous in the region where the
lower-bound condition is imposed.
Since $p_r=-\rho$ identically, the energy conditions take a particularly
simple form. The null energy condition, $\rho+p_i\ge0$ for every
principal pressure ($i=r,\theta,\varphi$), reads
\begin{equation}\label{eq:nec}
\rho+p_r=0,\qquad
\rho+p_\tau=\frac{2m'-r m''}{8\pi r^{2}}\ge0
\quad\Longleftrightarrow\quad
m''\le\frac{2m'}{r}.
\end{equation}
The strong energy condition, $\rho+p_r+2p_\tau\ge0$
together with $\rho+p_i\ge0$, under $\rho\geq0$ reads
\begin{equation}\label{eq:sec}
\rho+p_r+2p_\tau=2p_\tau=-\frac{m''}{4\pi r}\ge0
\quad\Longleftrightarrow\quad m''\le0.
\end{equation}

For later reference we collect the three standard energy conditions in terms
of the mass function. Under the standing assumption $\rho\geq0$, they read
\begin{equation}
\text{weak:}\quad m'\geq0,\ \ m''\leq\frac{2m'}{r};\qquad
\text{null:}\quad m''\leq\frac{2m'}{r};\qquad
\text{strong:}\quad m''\leq0.
\label{eq:ec}
\end{equation}
where $m'\geq0$ is displayed explicitly because it is retained throughout
this paper. The null energy condition by itself does not include $\rho\geq0$.
With the standing assumption, the weak and null energy conditions coincide,
while the strong energy condition implies both. We now show that none of these conditions
is sufficient to yield a positive lower bound on $T_{\min}/M$.

We now explain why the energy conditions in \eqref{eq:ec}, even when imposed
together, do not imply a universal positive lower bound on
$T_{\min}/M$. The period $T(u)$ at a given photon sphere is fixed by
the local geometry, whereas the ADM mass $M$ also includes matter at
arbitrarily large radii. We can therefore keep the geometry around
$u$ unchanged and modify the exterior matter distribution so that
$m'\geq0$ and $m''\leq0$. By choosing $m'$ to be positive and
constant over an arbitrarily distant interval of increasing length
and then smoothing it to zero over a fixed distance, we make the ADM
mass arbitrarily large while preserving $T(u)$ and the energy
conditions. The tail can be chosen with $m/r<1/2$, so that no new
outer horizon appears. Consequently, $T_{\min}/M\leq T(u)/M$ can be
made arbitrarily small, and no model-independent constant $c_0>0$
can satisfy $T_{\min}\geq c_0M$ for all spacetimes in this class.

Using the Einstein equations \eqref{eq:fluid}, we strengthen all three energy
conditions in \eqref{eq:ec} by considering the one-parameter family of
sufficient conditions
\begin{equation}
(\mathrm C_\kappa):\qquad
p_\tau\geq\frac{\kappa}{2}\rho,~~\rho\geq0;
\quad\Longleftrightarrow\quad
m''(r)+\kappa\frac{m'(r)}r\leq0,~~m'(r)\geq0.
\label{eq:Ckappa}
\end{equation}
The non-strict formulation also covers vacuum regions, where
$\rho=m'=0$. In particular, the Schwarzschild solution satisfies
$(\mathrm C_\kappa)$ for every $\kappa>0$.
For $\kappa>0$ and $m'\geq0$, condition $(C_{\kappa})$ implies $m''\leq0$ and
hence the full weak, null and strong energy conditions hold.
The usual charged black holes correspond to
$\kappa=2$: Reissner--Nordstr\"om saturates $(\mathrm C_2)$, i.e.,
$rm''+2m'=0$. We will derive a positive lower bound only
for $\kappa>1$.

We impose $(C_{\kappa})$ for all $r\geq r_{g}$, where $r_{g}$ is the largest
radius at which $T(r)$ attains its global minimum. Such a radius exists because
$T(r)\to+\infty$ as $r\to r_{h}^{+}$ and as $r\to\infty$, so the set of global
minimizers is a nonempty compact subset of $(r_{h},\infty)$ and hence has a
largest element. Note that $r_{g}$ is an interior critical point of $T(r)$, but
it need not coincide with the outermost critical point $u$.
Condition $(C_\kappa)$ gives
\begin{equation}
(r^{\kappa}m')'=r^{\kappa-1}(rm''+\kappa m')\le0,
\end{equation}
hence $r^\kappa m'(r)\leq r_g^\kappa m'(r_g)$ or
$m'(r)\leq m'(r_g)(r_g/r)^\kappa$ for $r\geq r_g$, and therefore
\begin{equation}
M=m(r_g)+\int_{r_g}^{\infty}m'(r)\,dr
 \leq m(r_g)+\int_{r_g}^{\infty}m'(r_g)
 \left(\frac{r_g}{r}\right)^\kappa dr
 =m(r_g)+\frac{m'(r_g)r_g}{\kappa-1},
 \label{eq:mass-est}
\end{equation}
provided $\kappa>1$. The actual ADM mass is already finite
by asymptotic flatness. The upper estimate for the remaining mass integral
converges precisely when $\kappa>1$. For $\kappa\leq1$, this particular
estimate diverges and provides no bound.

Since $r_g$ is a minimum of $T(r)$, one has $T'(r_g)=0$, which is
equivalent to the photon-sphere condition \eqref{eq:critical}. The
mass function therefore satisfies
\begin{equation}
m'(r_g)=\frac{3m(r_g)}{r_g}-1.
\label{ps}
\end{equation}

Combining \eqref{ps} with $m'(r_g)\ge0$ in \eqref{eq:Ckappa} and $f(r_g)=1-2m(r_g)/r_g>0$, we obtain
\begin{equation}
\mu=\frac{m(r_g)}{r_g}\in\left[\frac13,\frac12\right).
\end{equation}
The mass estimate \eqref{eq:mass-est} then becomes
\begin{equation}
M\le \mu r_g+\frac{(3\mu-1)r_g}{\kappa-1}
=r_g\frac{(\kappa+2)\mu-1}{\kappa-1}.
\end{equation}
Since $(\kappa+2)\mu-1>0$ for $\kappa>1$ and $\mu\ge1/3$, division by this
factor preserves the inequality direction. Evaluating \eqref{eq:T-general} at
$r_g$ gives $T_{\min}=2\pi r_g/\sqrt{1-2\mu}$, and hence
\begin{equation}
\frac{T_{\min}}{M}\ge
\frac{2\pi(\kappa-1)}{\sqrt{1-2\mu}\,[(\kappa+2)\mu-1]}.
\label{mu-bound}
\end{equation}

The right-hand side still involves the photon-sphere compactness
$\mu$, which varies from spacetime to spacetime.
The function $h(\mu)=\sqrt{1-2\mu}\,[(\kappa+2)\mu-1]$ is maximized at
$\mu_\ast=\dfrac{\kappa+3}{3(\kappa+2)}$
lying in $[1/3,1/2)$ for every $\kappa>1$, where
$h(\mu_\ast)=\dfrac{\kappa^{3/2}}{3\sqrt{3}\,\sqrt{\kappa+2}}$, and
substitution into \eqref{mu-bound} yields the bound:
\begin{equation}\label{lower-bound}
T_{\min}\ \ge\ c(\kappa)\,M,\qquad
c(\kappa)=6\sqrt{3}\,\pi\,
\frac{(\kappa-1)\sqrt{\kappa+2}}{\kappa^{3/2}}.
\end{equation}
Besides the three geometric assumptions inherited from Section~II, the
lower-bound proof requires the tangential-pressure condition $(C_{\kappa})$
with $\kappa>1$. This condition is sufficient for the estimate; it is not
asserted to be necessary for every possible positive lower bound.

The coefficient has the useful properties
\begin{equation}
\frac{d}{d\kappa}\log c(\kappa)
=\frac{3}{\kappa(\kappa-1)(\kappa+2)}>0,\qquad
\lim_{\kappa\to1^+}c(\kappa)=0,\qquad
\lim_{\kappa\to\infty}c(\kappa)=6\sqrt3\,\pi.
\end{equation}

Since $c(\kappa)$ is increasing, the strongest lower bound is obtained
by taking $\kappa$ as large as possible. For each model, the largest
allowed value is determined by the pointwise ratio
\begin{equation}
K(r)=-\frac{r m''(r)}{m'(r)}.
\end{equation}

Therefore, the largest value of $\kappa$ that can be used for the
whole exterior region is
\begin{equation}
\kappa_g=\min_{r\geq r_g}K(r).
\end{equation}
For a vacuum, $(\mathrm C_\kappa)$
holds for arbitrarily large $\kappa$ and hence $\kappa_g=\infty$.
A direct substitution gives
\begin{equation}
K_{\rm RN}(r)=2,\qquad
K_{\rm H}(r)=\frac{4(r^{3}-L^{2}M)}{r^{3}+2L^{2}M},\qquad
K_{\rm B}(r)=\frac{3r^{2}-2g^{2}}{r^{2}+g^{2}}.
\end{equation}
For the Reissner--Nordstr\"om model, $K_{\rm RN}=2$ is constant.
For the Hayward and Bardeen models, $K(r)$ is increasing for $r>0$.

The lower bound \eqref{lower-bound} is a statement about the global
minimum of $T(r)$ outside the horizon. Its proof requires the condition
$(C_\kappa)$ throughout the range $r\geq r_g$.
So the minimum of
$K(r)$ in this range is attained at $r_{g}$. Thus, in all three nonvacuum families, the minimum on $[r_g,\infty)$ is
obtained analytically at $r_g$:
\begin{equation}
\kappa_g=\min_{r\geq r_g}K(r)=K(r_g).
\end{equation}

$r_g$ is the radius at which $T(r)$ attains its
global minimum, $\mu=m(r_g)/r_g$, and
$T_{\min}/M=2\pi r_g/(M\sqrt{1-2\mu})$. We only consider black-hole
configurations, so the parameters are restricted by the existence of
an event horizon. For Reissner--Nordstr\"om this requires $Q\leq M$;
for Hayward and Bardeen, the critical values $L_c=4M/(3\sqrt3)$ and
$g=g_c$ are determined by the horizon condition. The numerical values
are collected in Table I.

\begin{table}[htbp]
\centering
\caption{Numerical verification for the standard models at extremal parameters
(which minimize $T_{\min}/M$ within each family). The last column evaluates
the universal lower-bound coefficient at $\kappa_g$.}
\begin{tabular}{|l|c|c|c|c|c|}
\hline
 & $r_g/M$ & $\mu$ & $T_{\min}/M$ & $\kappa_g$ & $c(\kappa_g)$ \\
\hline
Schwarzschild     & 3.0000 & 1/3    & 32.648 & $\infty$ (vacuum) & $6\sqrt3\pi$ (limit) \\
\hline
RN ($Q=M$)        & 2.0000 & 0.3750 & 25.133 & 2.000 (saturated) & 23.086 \\
\hline
Hayward ($L=L_c$) & 2.6524 & 0.3545 & 30.894 & 3.642 & 29.478 \\
\hline
Bardeen ($g=g_c$) & 2.3012 & 0.3706 & 28.425 & 2.497 & 26.266 \\
\hline
\end{tabular}
\end{table}

\begin{figure}[h]
\centering
\includegraphics[width=0.52\textwidth]{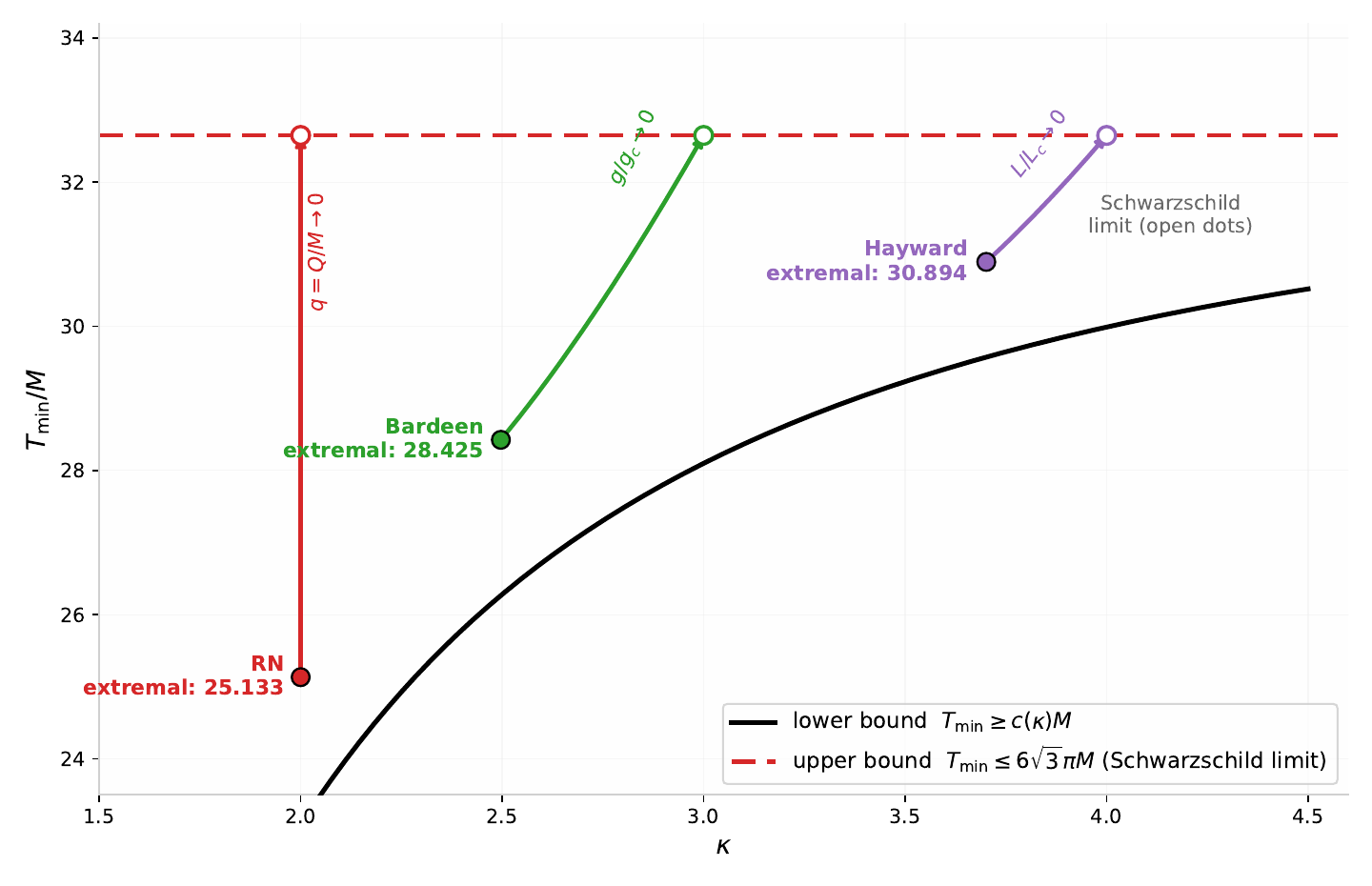}
\caption{The universal lower bound $T_{\min}\ge c(\kappa)M$
\eqref{lower-bound} (solid curve) with the standard black holes at
their own $\kappa_g$ and actual $T_{\min}/M$. All model
points lie on or above the curve, and the upper dashed line is the universal
upper bound saturated by Schwarzschild. The lower-bound coefficient is not
claimed to be sharp.}
\label{fig:models}
\end{figure}

In summary, the lower bound \eqref{lower-bound} gives a one-parameter family
$T_{\min}\geq c(\kappa)M$ with $\kappa>1$. We apply this bound to
each standard black hole with its own $\kappa_g$. In the
$(\kappa,T_{\min}/M)$ plane, each family traces a curve segment
between the lower-bound coefficient and the upper bound, and the
curve terminates on the upper bound in the Schwarzschild limit
in Fig.~2.

\section{Astrophysical implications}\label{sec:astro}

Translated into the frequency of the fastest light ring,
$\Omega_{*}\equiv2\pi/T_{\min}$, the two bounds confine every black
hole of the class satisfying $(\mathrm{C}_{\kappa})$ to the window
\begin{equation}
\frac{1}{3\sqrt{3}\,M}\ \leq\ \Omega_{*}\ \leq\
\frac{2\pi}{c(\kappa)\,M},
\end{equation}
whose edges stand in the ratio $6\sqrt{3}\pi/c(\kappa)$: for $\kappa=2$
the upper edge exceeds the lower one by a factor of exactly $\sqrt{2}$, i.e.
numerically $\Omega_{*}M\in[0.19245009,0.27216553]$, and the window collapses
to the single Schwarzschild frequency in the vacuum limit $\kappa\to\infty$.
Such light-ring frequencies underlie the characteristic photon-ring
features of black holes \cite{Cardoso2014,Hod4,Cardoso2016}.

\section{Conclusions}\label{sec:concl}

We have determined bounds on the minimum orbital period of four-dimensional
black holes. For static spherically symmetric black holes with an event
horizon, asymptotic flatness and non-negative energy density, the upper bound
$T_{\min}\le6\sqrt3\pi M$ is a theorem. The bound is sharp and is saturated by
the Schwarzschild solution.
A positive lower bound in terms of the total ADM mass requires additional
exterior control. We gave an asymptotically flat construction satisfying the
weak, null, and strong energy conditions for which $T_{\min}/M$ tends to zero.
These energy conditions alone therefore cannot imply a universal positive
lower bound.
Under the additional tangential-pressure condition $(C_\kappa)$, if
$p_\tau\geq(\kappa/2)\rho$ holds for some $\kappa>1$ and for all $r\geq r_g$,
then $T_{\min}\geq c(\kappa)M$.
The Schwarzschild, Reissner--Nordstr\"om,
Hayward and Bardeen black holes are covered by this bound, each with its own
$\kappa_g$. Open problems are to find broader exterior conditions that
still yield a universal positive lower bound.

\begin{acknowledgments}
This work was supported by the Shandong Provincial Natural Science Foundation of China under Grant
No.~ZR2022MA074. This work was also supported by a grant from Qufu Normal University of China.
\end{acknowledgments}

\end{document}